\documentclass[prl,twocolumn,showpacs]{revtex4}
\usepackage{graphicx}
\usepackage{bm}
\usepackage{amsmath}
\usepackage{amsfonts}
\usepackage{amssymb}

\newcommand{\refeq}[1]{(\ref{#1})}
\newcommand{\reffig}[1]{FIG. \ref{#1}}
\newcommand{\defeq}{\stackrel{\mathrm{def}}{=}}

\begin{document}
\title{
Collective motion in a Hamiltonian dynamical system
}
\author{Hidetoshi Morita}
\email{morita@complex.c.u-tokyo.ac.jp}
\altaffiliation[Present address:]{
Faculty of Science and Engineering,
Waseda University,
3-4-1 Okubo, Shinjuku-ku, Tokyo 169-8555, Japan
}
\author{Kunihiko Kaneko}
\affiliation{
Department of Basic Science,
Graduate School of Arts and Sciences,
The University of Tokyo,
3-8-1 Komaba, Meguro-ku, Tokyo 153-8902, Japan
}
\date{\today}

\begin{abstract}
Oscillation of macroscopic variables is discovered in a metastable state
in the Hamiltonian dynamical system of mean field XY model,
the duration of which is divergent with the system size.
This long-lasting periodic or quasiperiodic collective motion
appears through Hopf bifurcation,
which is a typical route in low-dimensional dissipative dynamical systems.
The origin of the oscillation is explained,
with self-consistent analysis of the distribution function,
as the emergence of self-excited ``swings'' through the mean-field.
The universality of the phenomena is also discussed.
\end{abstract}


\maketitle

Dissipative systems often show periodic, quasiperiodic,
and chaotic temporal behaviors at a macroscopic level,
when set at far away from equilibrium.
They are described as low-dimensional dynamics,
the discovery of which has marked an epoch
of nonlinear dynamics studies in physics.
Recalling that the involved degrees of freedom are large,
such macroscopic dynamics are a result of collective motion
out of high-dimensional microscopic dynamics.
Thus the collective motion is an important issue to be studied
from both viewpoints of dynamical systems with many degrees of freedom
and of nonequilibrium statistical mechanics.
Indeed it has been intensively and extensively studied
for systems consisting of a large number of chaotic elements
over one and a half decades
\cite{Kaneko1990,Chate_Manneville1992,col1,col2},
although the underlying microscopic dynamics in these studies
have been restricted to \textit{dissipative} chaos.

Does such collective motion exist in closed thermodynamic systems,
or Hamiltonian dynamical systems with many degrees of freedom?
According to the standard belief in statistical physics,
it seems rather difficult.
In equilibrium, by definition, it does not exist.
Also if the system rapidly relaxes to equilibrium,
there exists only decay process.
Thus long-term residence at a nonequilibrium state is at least required
for such collective motion.
Indeed, there have been some studies on long-lasting metastable state
in the relaxation to equilibrium, when some Hamiltonian dynamical systems
are set at certain initial conditions
\cite{Antoni_Ruffo1995,nega_spec_heat,Morita_Kaneko2004}.
However, no clear demonstration of
macroscopic low-dimensionality in such metastable state
has been reported yet, as far as the authors know.

Here we discover such collective oscillation
of macroscopic (thermodynamic) variables
in a closed Hamiltonian system with many degrees of freedom.
The oscillation is sustained over a long time,
and indeed the duration increases with the system size,
suggesting the divergence in the thermodynamic limit.
Furthermore, the oscillation appears in a similar way
to the bifurcation in low-dimensional dissipative dynamical systems,
suggesting the low-dimensionality of the collective motion.
In the present Letter, we report the essence of the phenomena,
while the details will be soon reported elsewhere~\cite{Fullpaper}.

We adopt the Hamiltonian dynamical system of mean field XY model,
or globally coupled pendula
\cite{Konishi_Kaneko1992,Inagaki1993,Antoni_Ruffo1995},
\begin{align}
{\cal H}=\sum_{i=1}^{N} \frac{p_i^2}{2}
+\frac{1}{2N}\sum_{i=1}^{N}\sum_{j=1}^{N} [1-\cos(\theta_i-\theta_j)].
\label{eq:Hamiltonian}
\end{align}
All the $N$ pendula interact with each other through phase difference.
Each pendulum has two types of motion;
rotation at a higher and libration at a lower energy.
The equilibrium state is exactly solvable
and uniquely determined only with the total energy density $U={\cal H}/N$.
Note that the system shows continuous phase transition.

We focus on the dynamics of the variance of momentum, $T(t)$,
and the magnitude of the mean field, $M(t)$, respectively defined as,
\begin{align}
T(t) \defeq \frac{1}{N}\sum_{j=0}^{N} p_j(t)^2,
\quad
M(t) e^{i\phi(t)} \defeq \frac{1}{N}\sum_{j=0}^{N} e^{i\theta_j(t)}.
\label{eq:macro}
\end{align}
These are macroscopic variables; indeed, they are nothing but
the temperature and the magnetization of the system in equilibrium.
Since \refeq{eq:Hamiltonian} yields the constraint,
\begin{equation}
2U=T(t)+1-M(t)^2,
\label{eq:UTM}
\end{equation}
we mainly discuss the dynamics of $M(t)$ in the following.
Using \refeq{eq:macro}, the equations of motion are described as
single pendula interacting with the mean field:
\begin{equation}
\dot{\theta_j}=p_j,\quad
\dot{p_j}=-M(t)\sin(\theta_j-\phi(t)).
\label{eq:Hamilton_eq}
\end{equation}

To study relaxation phenomena far from equilibrium,
we prepare nonequilibrium initial states.
In particular, we take the initial condition
of $\{\theta_i\}$ and $\{p_i\}$ given as follows.
For a given total energy density $U$,
assign an initial magnetization $M_0=M(0)$.
First, the initial distribution of $\{\theta_i\}$
is set as a Boltzmann distribution of the equilibrium state determined,
irrespective of $U$, with the magnetization $M_0$ and the temperature
$T_{eq}(M_0)$, where $T_{eq}(M)$ is the equation of state in equilibrium:
\begin{align}
f_\theta^0(\theta;M_0) = \frac{1}{Z_\theta(M_0)}
\exp\left[\frac{M_0}{T_{eq}(M_0)}\cos\theta\right],
\end{align}
where $Z_\theta(M_0)$ is a normalization.
Next, the distribution of $\{p_j\}$
is set as a Maxwell distribution determined with the temperature
$T_0(U,M_0)=2U-1+M_0^2$ to fulfill \refeq{eq:UTM}:
\begin{align}
f_p^0(p;U,M_0) = \frac{1}{Z_p(U,M_0)}
\exp\left[-\frac{p^2}{2T_0(U,M_0)}\right],
\end{align}
where $Z_p(U,M_0)$ is a normalization.
Here $\theta_i$ and $p_i$ are given independently of each other.
To sum up, the initial distribution is given as,
\begin{equation}
f_0(\theta,p;U,M_0)=f_\theta^0(\theta;M_0) f_p^0(p;U,M_0).
\end{equation}
The present nonequilibrium initial conditions are
controlled systematically with $(U,M_0)$.
Here, if $M_0$ is the equilibrium value for the given $U$,
then $f_0$ is exactly the equilibrium distribution.
In this sense, the present initial states are connected
to equilibrium on the $(U,M_0)$ plane.

\begin{figure}
\begin{center}
\includegraphics[width=0.4\textwidth]{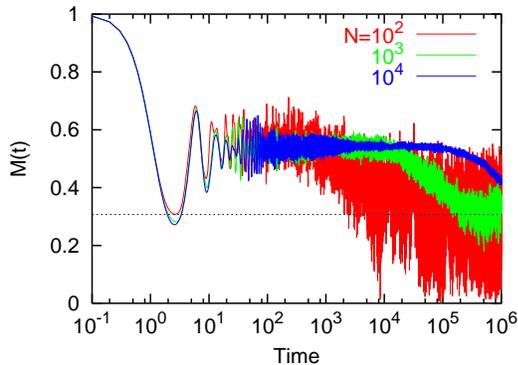}
\caption{
A typical time series of $M(t)$. $U=0.69$~\cite{nega_spec_heat}, $M_0=1$.
The dotted line indicates the equilibrium value.
The abscissa axis is log scale.
}
\label{fig:metasta}
\end{center}
\end{figure}

Now we study the relaxation to equilibrium.
A typical time series of $M(t)$ is shown in \reffig{fig:metasta}.
At the beginning ($t\gtrsim 0$), $M(t)$ decays almost exponentially.
However, $M(t)$, then, does not simply reach equilibrium
but stays at around a larger value than the equilibrium value
over a quite long time ($t\sim 10^4$).
After the long interval,
$M(t)$ departs from the plateau toward equilibrium logarithmically.
Finally, $M(t)$ reaches the equilibrium value, and fluctuates around there.
As shown, the duration time in the plateau increases with $N$, and indeed
for $N>10^5$, it is almost impossible to reach the equilibrium within 
numerical simulation (say $t<10^6$).
The result suggests that the metastable state is sustained
in a macroscopic time scale in the thermodynamic limit ($N\to\infty$).

Metastability in the present model has been intensively investigated
for a decade, especially by taking a rectangular (so
called \textit{waterbag}) initial momentum distribution
\cite{Antoni_Ruffo1995,nega_spec_heat}.
The metastable states exist only in the region
just below the critical energy of the phase transition,
and there $M(t)$ and $T(t)$ take smaller values than those in equilibrium.
This leads to a branch of negative specific heat,
which is discussed as a reflection of thermodynamic metastability,
i.e. the local minimum of the free energy potential.
On the other hand, the present metastable state takes
larger values of $M(t)$ and $T(t)$ than those in equilibrium,
and exists over a broader region across the critical point
than the negative specific heat branch.
Furthermore, since no local potential minimum exists around there,
the present metastable state is
not a thermodynamically but a dynamically stable state.
Thus the present metastability is a novel state.

\begin{figure}
\begin{center}
\includegraphics[width=0.4\textwidth]{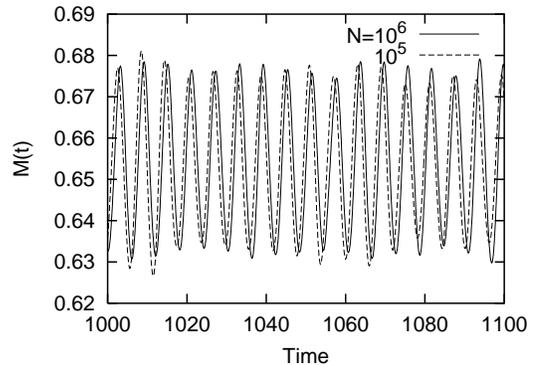}
\caption{A time series of $M(t)$ in the metastable state.
$N=10^6$ (solid) and $10^5$ (broken) for $U=0.5$ and $M_0=0.9$.
}
\label{fig:M_tseri}
\end{center}
\end{figure}

\begin{figure}
\begin{center}
\includegraphics[width=0.36\textwidth]{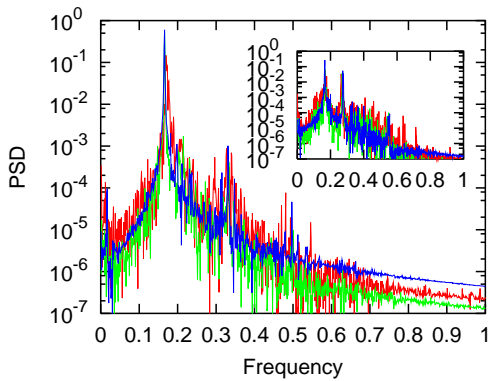}
\caption{The power spectrum density of $M(t)$ for $U=0.5$ and $0.75$ (inset).
$N=10^4$ (red), $10^5$ (green), and $10^6$ (blue). $M_0=0.9$.
}
\label{fig:powsp}
\end{center}
\end{figure}

We study the metastable state in more detail.
The closeup of the time series of $M(t)$ at the state
shows a periodic oscillation (\reffig{fig:M_tseri}).
The oscillation is not due to the finiteness of $N$.
In fact, the oscillation becomes more apparent with increasing $N$,
in strong contrast with the fluctuation around equilibrium that reduces to zero.
The corresponding power spectrum density (\reffig{fig:powsp})
shows a large peak ($f\approx 0.166$) with its harmonic component,
which remains sharp with increasing $N$.
This indicates that the periodic motion survives
even in the thermodynamic limit.
Although in much longer time scale
the amplitude of the oscillation decreases little by little,
the decline becomes more gradual with increasing $N$,
which implies that the periodic motion lasts permanently
in the thermodynamic limit.
Thus we have discovered the collective periodic motion.

Besides the periodic motion, the macroscopic variables
take various temporal pattern depending on the total energy.
In higher energy region, $M(t)$ shows more complex collective motion
than just a periodic oscillation.
Its power spectrum (the inset of \reffig{fig:powsp}) indicates
that it is a quasiperiodic motion on a $T^2$ torus.
In lower energy region, on the other hand,
the system rapidly relaxes to equilibrium,
without staying in any apparent metastable state,
and there $M(t)$ is almost stationary.

Strictly speaking, there exist some fluctuation
around each of the stationary, periodic, and quasiperiodic motion,
which remain finite even with increasing $N$.
In fact, in the periodic region,
the locus of the delay coordinate vector of $M(t)$
in the embedded phase space~\cite{embed}
is not completely one-dimensional
but has a finite width around the periodic motion.
Similarly, even in the seemingly stationary region,
the fluctuation around the fixed point does not decrease with $N$.
This residual fluctuation is regarded as a result of
high-dimensional collective motion~\cite{Fullpaper},
as is observed in dissipative systems~\cite{Kaneko1990}.

Except for the fluctuation,
the temporal behavior of the macroscopic variable changes as
(almost) stationary $\to$ (almost) periodic $\to$ (almost) quasiperiodic
with increasing $U$.
This is regarded as a ``bifurcation'' of the collective motion.
Here we note the similarity to the typical bifurcation
in low dimensional dissipative dynamical systems,
fixed point $\to$ limit cycle $\to$ torus, through Hopf bifurcations.
Thus the present bifurcation of the collective motion is basically described as
that of low dimensional dynamical systems, in particular by Hopf bifurcations.

\begin{figure}
\begin{center}
\includegraphics[width=0.4\textwidth]{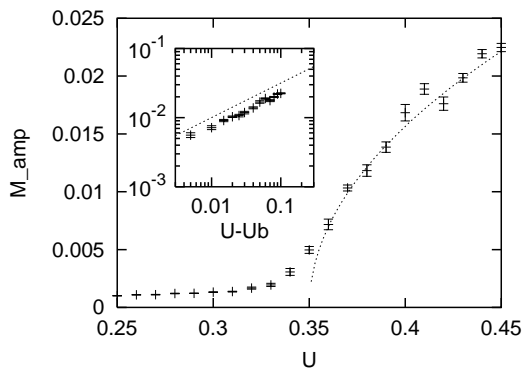}
\caption{
The amplitude of $M(t)$ against $U$,
in the vicinity of the bifurcation point.
$N=10^5$, $M_0=0.9$. 25 samples are calculated for each $U$.
The inset is the log-log plot of the same amplitude against $U-U_b$,
where $U_b=0.35$ is the ``bifurcation energy'' obtained from fitting.
The dotted lines indicate $M(t)\propto (U-U_b)^{1/2}$.
}
\label{fig:Hopf}
\end{center}
\end{figure}

We investigate the bifurcation in more detail.
We calculate the amplitude of $M(t)$ against $U$,
as shown in \reffig{fig:Hopf}.
The amplitude begins to increase above some critical energy $U_b$,
with the approximate dependence of $(U-U_b)^{1/2}$.
This suggests that the bifurcation from stationary
to periodic oscillation is a Hopf type.
We have also studied the change of the behavior
of the macroscopic variables against $M_0$,
which is another controllable for the initial distribution,
and confirmed again the Hopf-type bifurcation
from stationary to periodic oscillation.

\begin{figure}
\begin{center}
\includegraphics[width=0.39\textwidth]{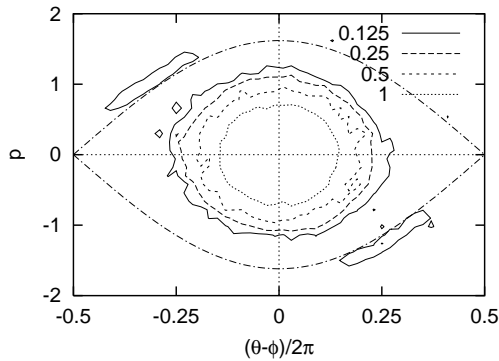}
\caption{
A snapshot of the contour of the one-body distribution function,
$f(\theta,p,t)$, corresponding to $t=1001.2$ of \reffig{fig:M_tseri}.
$N=10^5$. The dash-dotted line indicates the separatrix.
}
\label{fig:mu}
\end{center}
\end{figure}

Next, we study the origin of the collective motion,
in particular the periodic motion,
from the viewpoint of the microscopic dynamics.
Since it is almost impossible to investigate the trajectories
of such large degrees of freedom,
we study one-body distribution function, $f(\theta,p,t)$.
A snapshot of $f(\theta,p,t)$ for the collective periodic motion
when the phase of $M(t)$ is zero
(\reffig{fig:mu}) shows a pair of concentrated density
at $((\theta-\phi)/2\pi,p)\approx(\mp 0.3,\pm 1)$,
besides the expected high density at the center.
The pair of peaks rotates clockwise along the separatrix
keeping the localization, without diffusing out.
Since such localized rotation is absent in equilibrium,
it is important to ask its origin and relation to the collective motion.

To answer the question, we consider
an ensemble of $N$ independent parametrically driven pendula:
\begin{equation}
\dot{\theta_j}=p_j,\quad
\dot{p_j}=-M_{ext}(t)\sin\theta_j,
\label{eq:onebody}
\end{equation}
where $M_{ext}(t)=g+h\sin\Omega t$.
The equations are almost equal to \refeq{eq:Hamilton_eq}
except in that $M_{ext}(t)$ is externally applied,
while $M(t)$ is self-consistently determined in \refeq{eq:Hamilton_eq}.
Indeed, let $M_{int}(t)$ be the ``mean field'' given by \refeq{eq:macro},
and if $M_{int}(t)=M_{ext}(t)$, then the system is exactly
the original Hamiltonian system.
The Poincar\'{e} section of \refeq{eq:onebody}
on $\Omega t\equiv 0(\mod 2\pi)$,
an equivalent of the standard map~\cite{Chirikov1979},
yields islands of tori just at $(\theta/2\pi,p)\approx(\mp 0.3,\pm 1)$
that correspond to the 1:1 resonance on $\Omega$.
Moreover, even though we first distribute the elements homogeneously,
the axisymmetry is eventually broken down, while
the point symmetry is preserved due to the requirement of the dynamics;
for example, in a snapshot,
the first and third (the second and forth) quadrant are dense (sparse).

On the basis of the above analysis,
we give a self-consistent explanation on the collective periodic motion.
Once a considerable number of elements exist in the 1:1 island,
they make a periodic and finite contribution to $M(t)$.
Then the periodic $M(t)$ drives the elements,
which in turn makes the 1:1 island stable.
Moreover, the periodic drive also localizes other elements
by breaking the axisymmetry, which also stabilizes the periodicity.
Thus the periodic collective motion stably continues
with the self-organization of self-excited pendula, or ``swings.''

The above picture also explains the bifurcation
from the stationary to the periodic state as follows.
The 1:1 resonance state requires quite a little energy.
In low energy region, only a small number of elements can have enough energy to
stay in the 1:1 island, which does not contribute to the stability of $M(t)$.
In high energy region, on the other hand,
considerable elements become able to have enough energy,
which leads to stable periodic motion of $M(t)$.
Thereby, with increasing $U$, the bifurcation to periodic motion appears.
The bifurcation along the $M_0$-axis follows the same scenario;
as $M_0$ increases with constant $U$,
$T_0$ obtained from \refeq{eq:UTM} also increases,
whereby the elements have enough energy to stay in the 1:1 island.

In summary,
we have discovered collective periodic and quasiperiodic motion
at a macroscopic level
in the Hamiltonian dynamical system of mean field XY model.
In the thermodynamic limit, these motions are suggested to
continue forever, as their lifetime seems to diverge.
The collective motion appears through Hopf bifurcation,
which is a typical route in low-dimensional dissipative dynamical systems.
Thus we have observed macroscopic low-dimensional motion
out of microscopic high-dimensional Hamiltonian chaotic dynamics.
The mechanism of the macroscopic periodic motion is
explained from the viewpoint of the underlying microscopic dynamics,
as the self-organization of self-excited swings.

The above mechanism can work for Hamiltonian dynamical systems
with global coupling, or under mean field approximation,
when the resulting one-body dynamics have the separatrix of motion,
which allows for the self-consistent resonance of mean field oscillation.
The existence of such separatrix in the one-body dynamics means
the phase transition in terms of statistical physics.
Hence it is expected that the collective motion we discussed here
can be universally observed in a system
with relevant mean field interaction and phase transition.
In fact, we have recently discovered that the mean field $\phi^4$ model
does show collective periodic motion as well as Hopf bifurcation,
as in the present case~\cite{Fullpaper}.
Since systems with global or long-ranged coupling with
phase transition broadly exist in nature,
for example in molecular clusters and (bio)polymer systems,
it will be an important issue if the
collective motions are observed in such real systems.

In the end, we add a comment on Ref.~\cite{Dauxouis_Lepri_Ruffo2003}
that have reported ``coherent oscillating cluster''
in the metastable state of the negative specific heat branch
in the mean field $\phi^4$ model.
Although it may be a kind of collective motion,
it probably corresponds to the residual fluctuation
in the almost stationary motion
rather than the collective oscillation in our result,
since the oscillation amplitude is quite small.
Actually, in the present XY model, some tiny peaks in the power spectrum
are also observed at the stationary motion when $M_0=1$~\cite{Fullpaper}.
It is also noted that the Hopf bifurcation
as in low-dimensional dynamical system is confirmed in our case.
In fact, we have also discovered such low-dimensional collective motion
even in the mean field $\phi^4$ model,
with initial distribution as in the present Letter
rather than that adopted in Ref.~\cite{Dauxouis_Lepri_Ruffo2003}.
Finally, the present oscillation is observed,
not in the negative specific heat branch,
as is also in contrast with the gravothermal oscillation
\cite{Makino1996} in a self-gravitating system.

The authors are grateful to Kensuke Ikeda for discussion.
This work was supported by a Grant-in-Aid for Scientific Research
from MEXT Japan.


\begin{thebibliography}{19}
\expandafter\ifx\csname natexlab\endcsname\relax\def\natexlab#1{#1}\fi
\expandafter\ifx\csname bibnamefont\endcsname\relax
  \def\bibnamefont#1{#1}\fi
\expandafter\ifx\csname bibfnamefont\endcsname\relax
  \def\bibfnamefont#1{#1}\fi
\expandafter\ifx\csname citenamefont\endcsname\relax
  \def\citenamefont#1{#1}\fi
\expandafter\ifx\csname url\endcsname\relax
  \def\url#1{\texttt{#1}}\fi
\expandafter\ifx\csname urlprefix\endcsname\relax\def\urlprefix{URL }\fi
\providecommand{\bibinfo}[2]{#2}
\providecommand{\eprint}[2][]{\url{#2}}

\bibitem{Kaneko1990}
\bibinfo{author}{\bibfnamefont{K.}~\bibnamefont{Kaneko}},
  \bibinfo{journal}{Phys. Rev. Lett.} \textbf{\bibinfo{volume}{65}},
  \bibinfo{pages}{1391} (\bibinfo{year}{1990}),
  \bibinfo{journal}{Physica D} \textbf{\bibinfo{volume}{55}},
  \bibinfo{pages}{368} (\bibinfo{year}{1992}).

\bibitem{Chate_Manneville1992}
\bibinfo{author}{\bibfnamefont{H.}~\bibnamefont{Chat\'{e}}} \bibnamefont{and}
  \bibinfo{author}{\bibfnamefont{P.}~\bibnamefont{Manneville}},
  \bibinfo{journal}{Prog. Theor. Phys.} \textbf{\bibinfo{volume}{87}},
  \bibinfo{pages}{1} (\bibinfo{year}{1992}).

\bibitem{col1}
\bibinfo{author}{\bibfnamefont{G.}~\bibnamefont{Perez}} \bibnamefont{and}
  \bibinfo{author}{\bibfnamefont{H.~A.} \bibnamefont{Cerdeira}},
  \bibinfo{journal}{Phys. Rev. A} \textbf{\bibinfo{volume}{46}},
  \bibinfo{pages}{7492} (\bibinfo{year}{1992});
\bibinfo{author}{\bibfnamefont{N.}~\bibnamefont{Nakagawa}} \bibnamefont{and}
  \bibinfo{author}{\bibfnamefont{Y.}~\bibnamefont{Kuramoto}},
  \bibinfo{journal}{Prog. Theor. Phys.} \textbf{\bibinfo{volume}{89}},
  \bibinfo{pages}{313} (\bibinfo{year}{1993});
\bibinfo{author}{\bibfnamefont{A.~S.} \bibnamefont{Pikovsky}} \bibnamefont{and}
  \bibinfo{author}{\bibfnamefont{J.}~\bibnamefont{Kurths}},
  \bibinfo{journal}{Phys. Rev. Lett.} \textbf{\bibinfo{volume}{72}},
  \bibinfo{pages}{1644} (\bibinfo{year}{1994}).

\bibitem{col2}
\bibinfo{author}{\bibfnamefont{T.}~\bibnamefont{Shibata}} \bibnamefont{and}
  \bibinfo{author}{\bibfnamefont{K.}~\bibnamefont{Kaneko}},
  \bibinfo{journal}{Phys. Rev. Lett.} \textbf{\bibinfo{volume}{81}},
  \bibinfo{pages}{4116} (\bibinfo{year}{1998}{\natexlab{b}});
\bibinfo{author}{\bibfnamefont{T.}~\bibnamefont{Shibata}},
  \bibinfo{author}{\bibfnamefont{T.}~\bibnamefont{Chawanya}}, \bibnamefont{and}
  \bibinfo{author}{\bibfnamefont{K.}~\bibnamefont{Kaneko}},
  \bibinfo{journal}{Phys. Rev. Lett.} \textbf{\bibinfo{volume}{82}},
  \bibinfo{pages}{4424} (\bibinfo{year}{1999}).

\bibitem{Antoni_Ruffo1995}
\bibinfo{author}{\bibfnamefont{M.}~\bibnamefont{Antoni}} \bibnamefont{and}
  \bibinfo{author}{\bibfnamefont{S.}~\bibnamefont{Ruffo}},
  \bibinfo{journal}{Phys. Rev. E} \textbf{\bibinfo{volume}{52}},
  \bibinfo{pages}{2361} (\bibinfo{year}{1995}).

\bibitem{nega_spec_heat}
See, for example, recent papers;
  \bibinfo{author}{\bibfnamefont{V.}~\bibnamefont{Latora}}, \bibnamefont{and}
  \bibinfo{author}{\bibfnamefont{A.}~\bibnamefont{Rapisarda}},
  \bibinfo{journal}{Physica D} \textbf{\bibinfo{volume}{193}},
  \bibinfo{pages}{315} (\bibinfo{year}{2004});
  \bibinfo{author}{\bibfnamefont{Y.~Y.} \bibnamefont{Yamaguchi}}
  \bibnamefont{et~al.}, \bibinfo{journal}{Physica A}
  \textbf{\bibinfo{volume}{377}}, \bibinfo{pages}{36} (\bibinfo{year}{2004});
and references therein.

\bibitem{Morita_Kaneko2004}
\bibinfo{author}{\bibfnamefont{H.}~\bibnamefont{Morita}} \bibnamefont{and}
  \bibinfo{author}{\bibfnamefont{K.}~\bibnamefont{Kaneko}},
  \bibinfo{journal}{Europhys. Lett.} \textbf{\bibinfo{volume}{66}},
  \bibinfo{pages}{198} (\bibinfo{year}{2004}).

\bibitem{Fullpaper}
\bibinfo{author}{\bibfnamefont{H.}~\bibnamefont{Morita}} \bibnamefont{and}
  \bibinfo{author}{\bibfnamefont{K.}~\bibnamefont{Kaneko}}, \bibinfo{note}{in
 preparation.}

\bibitem{Konishi_Kaneko1992}
\bibinfo{author}{\bibfnamefont{T.}~\bibnamefont{Konishi}} \bibnamefont{and}
  \bibinfo{author}{\bibfnamefont{K.}~\bibnamefont{Kaneko}},
  \bibinfo{journal}{J. Phys. A} \textbf{\bibinfo{volume}{25}},
  \bibinfo{pages}{6283} (\bibinfo{year}{1992}).

\bibitem{Inagaki1993}
\bibinfo{author}{\bibfnamefont{S.}~\bibnamefont{Inagaki}},
  \bibinfo{journal}{Prog. Theor. Phys.} \textbf{\bibinfo{volume}{90}},
  \bibinfo{pages}{577} (\bibinfo{year}{1993}).

\bibitem{embed}
\bibinfo{author}{\bibfnamefont{N.~H.} \bibnamefont{Packard}}
  \bibnamefont{et~al.}, \bibinfo{journal}{Phys. Rev. Lett.}
  \textbf{\bibinfo{volume}{45}}, \bibinfo{pages}{712} (\bibinfo{year}{1980});
\bibinfo{author}{\bibfnamefont{F.}~\bibnamefont{Takens}},
  \emph{Detecting Strange Attractors in Turbulence},
  in \emph{\bibinfo{booktitle}{Lecture Notes in Mathematics}}
  (\bibinfo{publisher}{Springer, New York}, \bibinfo{year}{1981}), vol.
  \bibinfo{volume}{898}.

\bibitem{Chirikov1979}
\bibinfo{author}{\bibfnamefont{B.~V.} \bibnamefont{Chirikov}},
  \bibinfo{journal}{Phys. Rep.} \textbf{\bibinfo{volume}{52}},
  \bibinfo{pages}{263} (\bibinfo{year}{1979}).

\bibitem{Dauxouis_Lepri_Ruffo2003}
\bibinfo{author}{\bibfnamefont{T.}~\bibnamefont{Dauxois}},
  \bibinfo{author}{\bibfnamefont{S.}~\bibnamefont{Lepri}}, \bibnamefont{and}
  \bibinfo{author}{\bibfnamefont{S.}~\bibnamefont{Ruffo}},
  \bibinfo{journal}{Nonlin. Sci. Num. Sim.} \textbf{\bibinfo{volume}{8}},
  \bibinfo{pages}{375} (\bibinfo{year}{2003}).

\bibitem{Makino1996}
\bibinfo{author}{\bibfnamefont{J.}~\bibnamefont{Makino}},
  \emph{Gravothermal Oscillation},
  in \emph{\bibinfo{booktitle}{Dynamical Evolution of Star Clusters}},
  (\bibinfo{publisher}{Kluwer},
  \bibinfo{address}{Dordrecht}, \bibinfo{year}{1996}).

\end{thebibliography}
\end{document}